\documentclass{article}

\usepackage{epsfig}

\usepackage{amsfonts}
\usepackage{amssymb}
\usepackage{amsmath}
\usepackage{epsfig}
\usepackage{color}
\usepackage{graphicx}

\newtheorem{theorem}{Theorem}[section]
\newtheorem{corollary}[theorem]{Corollary}

\newtheorem{lemma}[theorem]{Lemma}
\newtheorem{claim}[theorem]{Claim}
\newtheorem{fact}[theorem]{Fact}

\newcommand{\ignore}[1]{ }

\newcommand{\sq}{\hbox{\rlap{$\sqcap$}$\sqcup$}}
\newcommand{\qed}{\hspace*{\fill}\sq}
\newenvironment{proof}{\noindent {\bf Proof.}\ }{\qed\par\vskip 4mm\par}

\def\rep{\mbox{rep}}
\def\relay{\mbox{relay}}

\def\RR{\mathbb{R}}
\def\ZZ{\mathbb{Z}}

\def\NN{\mbox{NN}}
\def\UDG{\mbox{UDG}}
\def\US{\mbox{UDG-SENS}}
\def\NS{\mbox{NN-SENS}}

\def\pr{\mbox{P}}

\def\ktiles{\mbox{k}}
\def\utiles{\mbox{u}}

\def\lambdas{1.568}

\def\loc{\mbox{\sf location}}
\def\id{\mbox{\sf id}}
\def\region{\mbox{\sf region}}

\begin{document}

\title{Sparse power-efficient topologies for wireless ad hoc sensor
networks\thanks{ This work has been submitted to the IEEE for possible
publication. Copyright may be transferred without notice, after
which this version may no longer be accessible.}}

\author{
 Amitabha Bagchi\\Dept of Computer Science and Engg\\ Indian Institute
 of Technology\\ Hauz Khas, New Delhi 110016, India.\\ {bagchi@cse.iitd.ernet.in.}
}
\maketitle 

%=========================================================================
%  Abstract
%=========================================================================

\begin{abstract}
We study the problem of power-efficient routing for multihop wireless
ad hoc sensor networks. The guiding insight of our work is that unlike
an ad hoc wireless network, a wireless ad hoc sensor network does not
require full connectivity among the nodes. As long as the sensing
region is well covered by connected nodes, the network can perform its
task. We consider two kinds of geometric random graphs as base
interconnection structures: unit disk graphs $\UDG(2,\lambda)$ and
$k$-nearest-neighbor graphs $\NN(2,k)$ built on points generated by a
Poisson point process of density $\lambda$ in $\RR^2$. We provide
subgraph constructions for these two models $\US(2,\lambda)$ and
$\NS(2,k)$ and show that there are values $\lambda_s$ and $k_s$ above
which these constructions have the following good properties: (i) they
are sparse; (ii) they are power-efficient in the sense that the graph
distance is no more than a constant times the Euclidean distance
between any pair of points; (iii) they cover the space well; (iv) the
subgraphs can be set up easily using local information at each
node. Our analyses proceed by coupling the random graph constructions
in $\RR^2$ with a site percolation process in $\ZZ^2$ and using the
properties of the latter to derive properties of the former.  An
important consequence of our constructions is that they provide new
upper bounds for the critical values of the parameters $\lambda$ and
$k$ for the models $\UDG(2,\lambda)$ and $\NN(2,k)$. We also describe
a simple local algorithm requiring only location information (from a
GPS for example) and communication with immediate neighbors for
setting up the subnetworks $\US(2,\lambda)$ and $\NN(2,k)$ and for
routing packets on them.
\end{abstract}

%=========================================================================
%  Introduction
%=========================================================================

\section{Introduction}
\label{sec:intro}

Multihop sensor networks, where nodes act not only to sense but also
to relay information, have proven advantages in terms of energy
efficiency over single hop sensor networks~\cite{karl:2005}. Not only
is sensor-to-sensor communication useful for necessary tasks like time
synchronization~\cite{vangreunen-wsna:2003}, for certain kinds of
collaborative sensing functions like target
tracking~\cite{zhao-ieee:2003} sensor-to-sensor communication is
essential. The question of how to achieve connectivity arises here
just as it does in ad hoc wireless networks with one crucial
difference:
\begin{quote}
It is not necessary that every sensor be part of a connected network.
It is only necessary that the density of connected sensors is high
enough to perform the sensing function.
\end{quote}
In other words, even if some sensors are wasted in the sense that the
data they sense cannot be relayed, it does not matter as long as the
area being sensed is well covered with useful sensors which are part
of multihop network that can relay data. The difference from other ad
hoc wireless networks is that each node {\em expects} connectivity as
a service provided to it, while in the WASN individual nodes are not
important, the overall task is. In this paper we follow this critical
insight to propose sparse easy-to-compute power-efficient
constructions for multihop WASNs.

We consider two different types of geometric random graphs as the base
interconnection structures.  The nodes are modelled by a Poisson point
process with density $\lambda$ on the plane, $\RR^2$. The
interconnections between these points are modeled in two ways: 1) {\em
Unit Disk Graphs} in which there is an edge between two nodes if the
Euclidean distance between them is at most 1 and 2) {\em $k$-nearest
neighbor graphs} in which each node establishes (undirected) edges to
the $k$ points nearest to it. We will refer to the former model as
$\UDG(2,\lambda)$ and the latter as $\NN(2,k)$ (following the notation
introduced in~\cite{haggstrom-rsa:1996}). The 2 in both these terms
denotes the dimension of the space. Both these models display a
critical phenomenon. For $\UDG(2,\lambda)$ it is known that there is a
value $\lambda_c^{(2)}$ such that if $\lambda > \lambda^{(2)}_c$ the
graph contains an infinite component~\cite{meester:1996}. For
$\NN(2,k)$ the density is not relevant, instead Haggstr\"om and
Meester show that $k$ is the critical parameter
(see~\cite{haggstrom-rsa:1996}) i.e. there is a $k_c(2)$ such that for
all $k> k_c(2)$, $\NN(2,k)$ has an infinite component. We show that as
long as the critical parameters take at least certain values (which
are higher than the critical values) it is possible in both cases to
construct a subgraph of the infinite component with the following
properties:
\begin{description}
\item{P1. ({\em Sparsity})} The subgraph has a maximum degree of 4.
\item{P2. ({\em Constant stretch})} The distance between any two points in
  the subgraph is at most a constant factor greater than the Euclidean
  distance between the points. 
\item{P3. ({\em Coverage})} The subgraph is infinite and the probability of
  a square region of $\RR^2$ {\em not} containing any points of the
  subgraph decays exponentially with the size of the region.
\item{P4. ({\em Local computability})} Each node can determine if it is part
  of the subgraph by using its location information and by
  communicating with its immediate neighbours.
\end{description}
The subgraph constructed for $\UDG(2,\lambda)$ is called
$\US(2,\lambda)$ and that constructed for $\NN(2,k)$ is called
$\NS(2,k)$. We will show that there is a value $\lambda_s$ (and $k_s$)
such that for all $\lambda \geq \lambda_s$ (resp. $k \geq k_s$),
$\US(2,\lambda)$ (resp. $\NS(2,k)$) has properties (P1)-(P4).  These
properties align well with the properties of power-efficient spanners
studied in the context of ad hoc wireless networks (see~\cite[pp
177-178]{santi-acmcs:2005}.) The difference being that not every point
of the point process is required to be part of the network as long as
the sensing function is satisfied (which the coverage property (P3)
ensures.).

Property (P2) is of major consequence to the power consumption of the
network. This follows from the relationship between the stretch in
distance between two points and the consequent increase in the power
consumed in communicating between them. Formally, if we consider a
wireless network $G$ formed on a set of nodes $V$ in $\RR^2$, the {\em
distance stretch}, $\delta$, of a subgraph $H \subseteq G$ is defined
as
\[ \delta = \max_{u,v \in V} \frac{d_H(u,v)}{d_G(u,v)},\]
where $d_G(u,v)$ is the graph distance between $u$ and $v$ in $G$ i.e.
distance between $u$ and $v$ using the edges graph $G$ and $d_H(u,v)$
is the graph distance in $H$.  Li, Wan and Wang~\cite[Lemma
2]{li-ccn:2001} showed that given a connection network $G$ and a
subgraph $H$ with distance stretch $\delta$, the power taken to
communicate between any two nodes is at most $\delta^\beta$ where
$\beta$ is a parameter varying between 2 and 5. Clearly a network with
property (P2) achieves a constant power stretch since the Euclidean
distance between two points is a lower bound on the distance between
them in both $\UDG(2,\lambda)$ and $\NN(2,k)$. Hence we claim that our
constructions are power-efficient up to a constant
factor. Additionally we have property (P3) that guarantees coverage of
the region being sensed. We show that the probability that a region
does not contain a point of $\US(2,\lambda)$ (or $\NS(2,k)$) decays
exponentially with the size of the region when $\lambda \geq
\lambda_s$ (respectively $k \geq k_s$). In both cases the decay is
sharper if a larger value of $\lambda$ is chosen. This allows us to
achieve a target coverage by increasing the density to a high enough
level.

The basic idea behind our constructions is to couple the random graph
in $\RR^2$ with a discrete site percolation process in $\ZZ^2$. The
subgraph we construct mimics the nodes of a percolated mesh. One
important byproduct of our constructions is that they give the best
known upper bounds on the critical values for both our setting. We
improve the best known bound of 213 (due to Teng and
Yao~\cite{teng-algorithmica:2007}) for the critical value of $k$ for
$\NN(2,k)$ to 188. We also improve the bound for the critical value of
$\lambda$ in $\UDG(2,\lambda)$ to $\lambdas$.

The construction is easy to realize using location information (which
can be obtained using a GPS) and local computation, hence satisfying
property (P4).  The algorithm for routing on our subgraph
constructions is based on a simple distributed algorithm for routing
on the percolated lattice given by Angel
et. al.~\cite{angel-podc:2005}.

In the rest of this section we introduce some notation and definitions
that will be required. We also discuss the various strands of research
relevant to our paper. In Section~\ref{sec:coupling} we describe our
constructions $\US(2,\lambda)$ and $\NN(2,k)$ and give lower bounds on
the values $\lambda_s$ and $k_s$ above which properties (P1)-(P4)
hold. The results regarding stretch and coverage are detailed in
\ref{sec:stretch}. In Section~\ref{sec:algorithmic} we discuss the
algorithmic issues involved in forming our subgraphs from the
underlying structure and also describe how to route packets once the
structures are made.

\subsection{Preliminaries}
\label{sec:intro:definitions}

We will use the notation $d(x,y)$ to denote the Euclidean distance
between two points $x,y \in \RR^2$. In general we will denote the
graph distance (i.e. the shortest path along the edges of the graph)
of two vertices $u,v$ of a graph $G$ by $d_G(u,v)$. For the random
graphs $\NN(2,k)$ and $\UDG(2,\lambda)$ we will use the notation
$d_n(u,v)$ and $d_u(u,v)$ to denote the graph distance between them. 

\paragraph{Poisson point processes} Our random point sets are
generated by homogenous Poisson point processes of intensity $\lambda$
in $\RR^2$. Under this model the number of points in a
region is a random variable that depends only on its $d$-dimensional
volume i.e. the number of points in a bounded, measurable set $A$ is
Poisson distributed with mean $\lambda V(A)$ where $V(A)$ is the
$d$-dimensional volume of A. Further, the random variables associated
with the number of points in disjoint sets are independent.

\paragraph{Unit disk graphs} The random graph model
$\UDG(2,\lambda)$ is defined as follows: Given a set of points $S$
generated by a Poisson point process in $\RR^2$ with density
$\lambda$, there is an edge between points $x \in S$ and $y \in S$ if
$d(x,y) \leq 1$.

\paragraph{$k$-nearest neighbor graphs} The random graph model
$\NN(2,k)$ defined as follows: Given a set of points $S$ generated by
a Poisson point process in $\RR^2$ there is an (undirected) edge
between points $x \in S$ the $k$ points in $S \setminus \{x\}$ that
are closest to $x$. Note that the event that two points have exactly
the same distance from a point is a measure 0 event, but for practical
purposes we can use any tie-breaking mechanism we deem fit.

\paragraph{Site percolation} Consider an infinite graph defined on
the vertex set $\ZZ^d$ with edges between points $x$ and $y$ such that
$\| x - y\|_1 = 1$. Site percolation is a probabilistic process on
this graph. Each point of $\ZZ^d$ is taken to be {\em open} with
probability $p$ and {\em closed} with probability $1 - p$. Hence we
have a sample space $\Omega = \prod_{x \in \ZZ^d}\{0,1\}$, individual
elements of which are {\em configurations} $\omega = (\omega(x) : x
\in \ZZ^d)$. The product of all the measures for individual points
forms a measure for the space of possible configurations. An edge
between two open vertices is considered open. All other edges are
considered closed. A component in which open vertices are connected
through paths of open edges is known as an open cluster. It is known
that there is a value $p_c$ such that for all $p > p_c$ the graph
obtained has an infinite open cluster. This value is known as the
critical probability. When $p > p_c$ then each point of $\ZZ^d$ has
some non-zero probability of being part of an infinite cluster.
\ignore{ An important tool in the study of percolation is the FKG
inequality. In order to state it we need to observe that the set of
configurations of a site percolation process has a natural partial
order defined on it. We say that $\omega_1 \preceq \omega_2$, for any
$\omega_1, \omega_2 \in \Omega$ if
\[\forall x \in \ZZ^d : \omega_1(x) \leq \omega_2(x).\]
An event $A$ is called {\em increasing} if the indicator variable
$I_A$ has the property that $I_A(\omega) \leq I_A(\omega')$ whenever
$\omega \preceq \omega'$. Increasing events have the following useful
property 
\begin{lemma}
\label{lem:fkg}
If $A$ and $B$ are increasing events 
\[\pr(A\cap B) \geq \pr(A)\cdot \pr(B).\]
\end{lemma}
This result can easily be extended to any countable number of events.
}
The reader is referred to~\cite{grimmett:1999} for a full treatment of
percolation and to \cite{bollobas:2006} for a recent update on some
new directions in this area.

When the lattice undergoes percolation, the path between two connected
vertices might become long and tortuous. We introduce some notation
for this setting. The distance between two lattice points $x,y \in
\ZZ^d$ will be denoted $D(x,y)$. When the lattice has been percoalted
with probability $p$, the distance will be denoted $D^p(x,y)$.  Antal
and Pisztora studied this setting and proved a powerful theorem which
we state here as a lemma~\cite[Theorems 1.1 and
1.2]{antal-ap:1996}. We adopt the restatement of Angel
et. al.~\cite[Lemma 8]{angel-podc:2005}.
\begin{lemma}
\label{lem:antal}
\cite{antal-ap:1996,angel-podc:2005} For any $p > p_c$ and any
  $x,y$ connected through an open path in a cube $M^d$ of the infinite
  lattice. For some $\rho, c_2 > 0$ depending
  only on the dimension and $p$ and for any $a > \rho \cdot
  D(x,y)$
\[pr(D^p(x,y) > a)) < e^{-c_2 a}.\]
\end{lemma}

\subsection{Related work}
\label{sec:intro:related}

\paragraph{Wireless networks} Topology control in wireless networks
has been studied extensively (see e.g. the surveys by
Santi~\cite{santi-acmcs:2005} and
Rajaraman~\cite{rajaraman-sigact:2002}.) Two important goals of the
research in this area have been ensuring connectivity of all nodes and
energy-efficiency. 

The approach has been to take the underlying topology as a unit disk
graph~\cite{sen-wn:1997} or a proximity graph (for which several
proposals exist c.f. surveys cited above) on a point set and then
construct some kind of spanning subgraph of this point set with low
degree, constant stretch and the property that each node can compute
its connections using local information.  Although this line of
research has the same flavour as our work, it is different in a
fundamental way - we do not require all nodes to be connected - and so
we do not survey the literature in detail instead referring the reader
to general surveys on topology control by and and a specific survey on
spanners by Li and Wang~\cite{li-handbook:2007}.

\paragraph{Geometric random graphs and percolation} The study of
  random graphs obtained by applying connection rules on stationary
point processes is known as continuum percolation. Meester and Roy's
monograph on the subject provides an excellent view of the deep theory
that has been developed around this general
setting~\cite{meester:1996}. $\UDG(2,\lambda)$ is studied
in~\cite{meester:1996}, where the existence and non-triviality of the
critical density is demonstrated. Kong and
Zeh~\cite{kong-infocom:2008} show a lower bound of 0.7698 on
$\lambda_c$. An upper bound of 3.372 was earlier shown by
Hall~\cite{hall-ap:1985}. Hall's paper states an upper bound 0.843 for
a model of intersecting spheres which scales by a factor of 4 using
scaling property of coninumm percolation models~\cite{meester:1996}.

The $\NN(d,k)$ model was introduced by H\"aggstr\"om and
Meester~\cite{haggstrom-rsa:1996}. They showed that there was a finite
critical value, $k_c(d)$ for all $d \geq 2$ such that an infinite
cluster exists in this model. They proved that the infinite cluster
was unique and that there was a value $d_0$ such that $k_c(d) = 2$ for
all $d > d_0$. Teng and Yao gave an upper bound of 213 for
$k_c(d)$~\cite{teng-algorithmica:2007}.

$k$-nearest neighbor graphs on random point sets contained inside a
finite region have been extensively studied. The major concern,
different from ours, has been to ensure that {\em all} the points
within the region are connected within the same cluster. Ballister,
Bollob\'as, Sarkar and Walters~\cite{ballister-aap:2005} showed that
the smallest value of $k$ that will ensure connectivity lies between
$0.3043 \log n $ and $0.5139 \log n$, improving earlier results of Xue
and Kumar~\cite{xue-wn:2004}. Ballister et. al. also studied the
problem of covering the region with the discs containing the
$k$-nearest neighbours of the points. We refer the reader
to~\cite{ballister-aap:2005} for an interesting discussion relating
this setting to earlier work by Penrose and others.

%=========================================================================
%  Couplings
%=========================================================================

\section{The subgraph constructions}
\label{sec:coupling}

\begin{figure}[htbp]
\begin{center}
\includegraphics{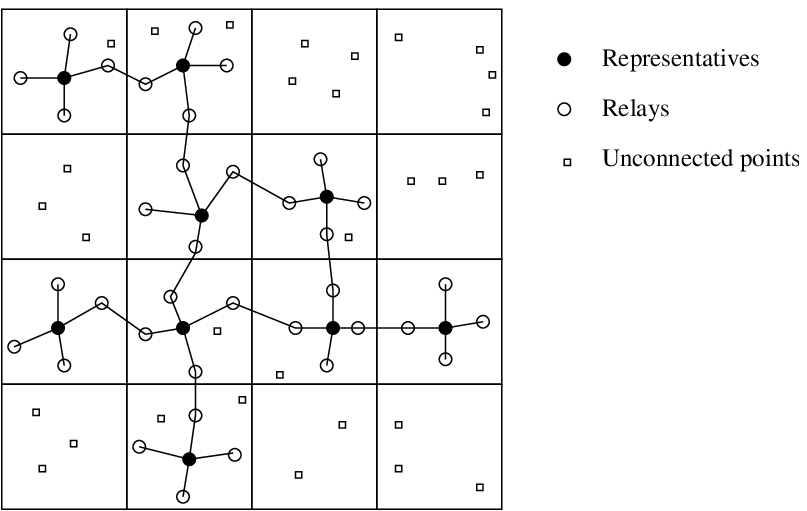}
\caption{A portion of the tiling of $\RR^2$.}
\label{fig:tiled-subgraph}
\end{center}
\end{figure}

In this section we describe our constructions and prove some important
properties. We begin by giving a general overview of our technique,
then move on to the specifics of the two settings.  For both
constructions we proceed by viewing $\RR^2$ as a union of a countably
infinite set of square tiles. Inside each tile we look for a two kinds
of points. The first kind is what we call a {\em representative
point.} Representative points lie at the centre of the tile, roughly
speaking. We also look for {\em relay points}, which help connect
representative points to neighbouring tiles. Both these kinds of
points have precise definitions that differ for $\US(2,\lambda)$ and
$\NS(2,k)$, we will discuss those in Section~\ref{sec:coupling:udg}
and~\ref{sec:coupling:nnk} respectively. A tile in which we find both
kinds of points we call a {\em good} tile, other tiles are {\em
bad}. We connect representative points to four relay points, one for
each neighbouring tile. Several points within and outside good tiles
may be left unconnected.  See Figure~\ref{fig:tiled-subgraph} for a
pictorial depiction. Note that representative points have degree 4 and
relay points have degree 2.

\begin{figure}[htbp]
\begin{center}
\includegraphics{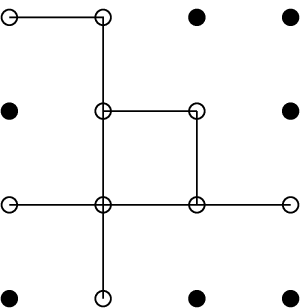}
\caption{The part of $\ZZ^2$ corresponding to the tiles shown in
  Figure~\ref{fig:tiled-subgraph}.} 
\label{fig:tiled-coupling}
\end{center}
\end{figure}

The subgraph drawn by connecting representative points through relay
points will be the network we will use for sensing. In order to prove
properties (P1)-(P4) we will couple the tiling with a site percolation
process in $\ZZ^2$. We associate each tile in $\RR^2$ with a point in
$\ZZ^2$. We declare a site in $\ZZ^2$ open only if the tile
corresponding to it in $\RR^2$ is good. Hence the probability that a
site is open is equal to the probability that its corresponding tile
is good. Our definitions of representative and relay points will
ensure that if two neighbouring tiles are good then their
representative points are connected through their relay points as
shown in Figure~\ref{fig:tiled-coupling}. This corresponds to the edge
between two open sites being open in $\ZZ^2$ (see
Figure~\ref{fig:tiled-coupling}.)

Since paths in $\ZZ^2$ corresponds to paths between points in $\RR^2$
it follows that if percolation occurs in the $\ZZ^2$ then an infinite
component must exist in the geometric random graph model as
well. Hence we can conclude that if the probability of a tile being
good exceeds the critical probability for site percolation, the
geometric random graph model also has an infinite component in it
almost surely. Let us now take a more specific look at the
constructions for $\UDG(2,\lambda)$ and $\NN(2,k)$.
\subsection{Unit-disk graphs}
\label{sec:coupling:udg}

\begin{figure}[htbp]
\begin{center}
\input{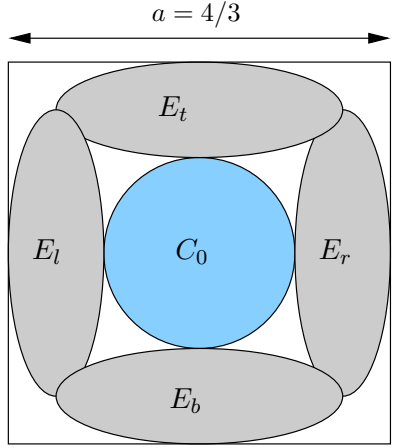}
\caption{A tile $t$ and it's 5 relevant regions. Note that the region
  $E_r$ is the part of the intersection lying wholly within $t$ of all
  unit discs centred at points in $C_0$ and in the region $E_l$
  (i.e. the left relay region) of the neighbour tile $t_r$.}
\label{fig:udg-coupling}
\end{center}
\end{figure}

The internal structure of a tile for the construction of $\US(2,\lambda)$ is
shown in Figure~\ref{fig:udg-coupling}.

We consider square tiles of side 4/3. For the sake of exposition let
us assume that the tile shown in Figure~\ref{fig:udg-coupling} is
centred at (0,0) and it's lower left corner is (-2/3, -2/3). Within
each tile we consider five disjoint regions, the representative region
$C_0(t)$, and the relay regions $E_l(t), E_r(t), E_t(t)$ and
$E_b(t)$. $C_0(t)$ is a circle of radius 1/2 centred at the
origin. The regions $E_i(t), i\in \{l, r, t, b\}$ are intuitive to
understand but slightly tricky to describe formally. We describe one
of them, $E_r(t)$. In order to do so, let us denote by $t_r$ the tile
immediately to the right of the tile $t$ i.e. the tile centred at
(4/3,0) with bottom left corner (2/3,-2/3). It's leftmost relay region
is $E_l(t_r)$. Now we define $E_r(t)$ as the part of the intersection
lying wholly within $t$ of all circles of unit radius centred at
points in $C_0(t)$ and $E_l(t_r)$. From this set we remove all the
points of $C_0(t)$. In the figure the region is depicted by an
ellipse, but clearly it is a less regular shape.

We call a tile $t$ {\em good} if each of $C_0(t)$ and $E_i(t), t \in
\{l, r, b, t\}$ contains at least one point of the point process. One
of the points contained in $C_0(t)$ will be the representative point
for this tile, denoted $\rep(t)$. Four other points, one from each of
the regions $E_i(t), t \in \{l, r, b, t\}$ will be the relay points
for this good tile. If a region has more than one point, the tie has
to be broken. This will be done in a distributed fashion. We postpone
the discussion of this aspect to Section~\ref{sec:algorithmic}. Note
that some of the relay regions overlap and hence it may be the case
that one point fulfils two relay functions.

According to the program described earlier we create a bijection,
$\phi$, between the tiles in $\RR^2$ and points in $\ZZ^2$ such that
neighbouring tiles in $\RR^2$ correspond to neighbouring points in
$\ZZ^2$. We couple $\UDG(2,\lambda)$ to a site percolation process in
$\ZZ^2$ by saying that a given point $x$ in $\ZZ^2$ is open only if
the tile $t = \phi^{-1}(x)$ is good. Now we can claim that the
existence of an edge in $\ZZ^2$ implies the existence of a path from
the representative points of the two tiles corresponding to the two
end points of the edge. We state this formally, including an
observation about the distance stretch between the two representative
points.

\begin{claim}
\label{clm:coupling:udg}
If an edge exists in the percolated mesh $\ZZ^2$ between two points
$x$ and $y$  then
\begin{enumerate}
\item There is a path between the representative points
  $\rep(\phi^{-1}(x))$ and $\rep(\phi^{-1}(y))$ of the tiles
  corresponding to $x$ and $y$ in $\UDG(2,\lambda)$ and
\item there is a constant $c_{\utiles} \leq 3$ such that
\begin{multline}
d_k(\rep(\phi^{-1}(x)), \rep(\phi^{-1}(y))) \\ \leq c_{\utiles} \cdot
d(\rep(\phi^{-1}(x)), \rep(\phi^{-1}(y))).
\end{multline}
\end{enumerate}
\end{claim}

\begin{figure}[htbp]
\begin{center}
\includegraphics{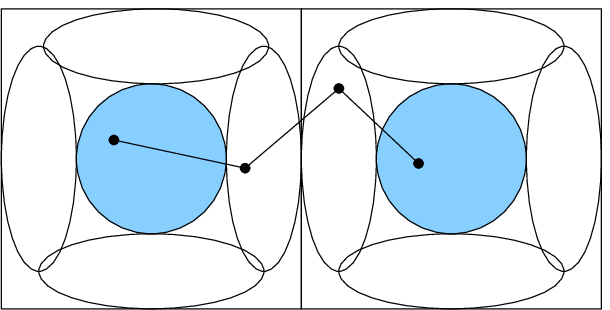}
\caption{A path between the representative points of neighboring good tiles.} 
\label{fig:udg-coupling-claim}
\end{center}
\end{figure}

\begin{proof}
Clearly if two neighbouring tiles $t$ and $t'$ are good, by the
goodness condition there will be an edge from the representative point
of one of them to a relay point in the direction of its neighbor. This
relay point will subsequently connect to the relay point of that
neighbor closest to it, which will in turn be connected to the
representative point of the neighbour (see
Figure~\ref{fig:udg-coupling-claim}). Clearly each of the three edges
on the path from $\rep(t)$ to $\rep(t')$ is at most 1 unit in length
so $c_{\utiles} \leq 3$.
\end{proof}

The largest connected component formed by the representative points
and relay points is $\US(2,\lambda)$. 

From Claim~\ref{clm:coupling:udg}, it is easy to deduce that if an
infinite component exists in the site percolation setting, then an
infinite component exists in $\UDG(2,\lambda)$. Hence we need to
determine for what values of $\lambda$ the site percolation process is
supercritical.  The critical probability for site percolation lies
between 0.592 and 0.593 (see e.g. \cite{lee-arxiv:2007}). Numerical
calculations showed that the smallest value of $\lambda$ for which the
probability of a tile being good exceeds 0.593 is $\lambda_s =
\lambdas$. Hence for $\lambda$ larger than this value $\US(2,\lambda)$
is infinite.

Since this improves the best known upper bound of
3.372~\cite{kong-infocom:2008}, we state it here as a theorem. 

\begin{theorem}
\label{thm:2d-bound-udg}
For $\UDG(2,\lambda)$
\[\lambda^{(2)}_c < \lambdas.\]
\end{theorem}

In Section~\ref{sec:stretch} we will show that $\US(2,\lambda)$ for
$\lambda \geq \lambda_s$ has constant stretch. For now we move on to
the construction of $\NS(2,k)$.

\subsection{Nearest-neighbor graphs}
\label{sec:coupling:nnk}

\begin{figure}[htbp]
\begin{center}
\input{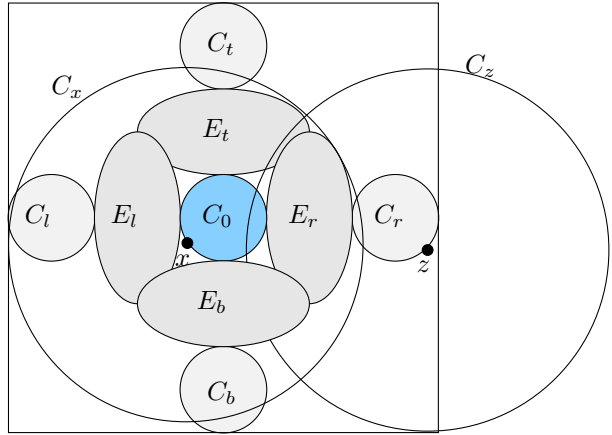}
\caption{A tile $t$ and it's 9 relevant regions. Note that the region
  $E_r$ lies wholly within all discs of the form $C_x$ and $C_z$
  centred at points on the boundary of the discs $C_0$ and $C_r$.}
\label{fig:2d-bound}
\end{center}
\end{figure}

The internal structure of a tile for the construction of $\NS(2,k)$ is
shown in Figure~\ref{fig:2d-bound}.

Let us say that the tile in Figure~\ref{fig:2d-bound} is centred at
$(0,0)$ with bottom left corner $(-5a,-5a)$ and top right corner
$(5a,5a)$. For convenience we will refer to the tiles surrounding the
tile $t$ as, couunterclockwise starting from the right $t_r$, $t_t$,
$t_l$ and $t_b$. We consider five circles of radius $a$: $C_0$ centred
at $(0,0)$, $C_l$ centred at $(-4a,0)$, $C_r$ centred at $(4a,0)$,
$C_t$ centred at $(0,4a)$ and $C_b$ centred at $(0,-4a)$. There are
four other region which are named $E_l, E_r, E_t$ and $E_b$ in the
figure. $E_r$ is defined as follows. Consider the largest circle
centred at any point in $C_0$ or $C_r$ that lies wholly within the two
tiles $t$ and $t_r$. Two such circles, $C_x$ and $C_z$, are depicted
in Figure~\ref{fig:2d-bound}. $E_r$ is the locus of the points
contained in all such circles. The regions $E_l, E_t$ and $E_b$ are
defined similary by $C_0$ alongwith $C_l, C_t$ and $C_b$ respectively
and the tiles $t_l, t_t$ and $t_b$ respectively.

Now, we call tile $t$ {\em good} if
\begin{enumerate}
\item the number of points inside $t$ is at most $k/2$ and 
\item the nine regions $C_0, C_r, C_t, C_l, C_b, E_r, E_t, E_l$ and
  $E_b$ contain at least one point each.
\end{enumerate}

One point contained in $C_0$ will be the representative point of the
tile $t$, denoted $\rep(t)$. A point from each of the other 8 regions
will be relay points. If these regions contain multiple points we will
have to select one from each and discard the rest. As in the case of
$\US(2,\lambda)$ we postpone the discussion of how to select this one
point each to Section~\ref{sec:algorithmic}.

According to the program described earlier we create a bijection,
$\phi$, between the tiles in $\RR^2$ and points in $\ZZ^2$ such that
neighbouring tiles in $\RR^2$ correspond to neighbouring points in
$\ZZ^2$. We couple $\NN(2,k)$ to a site percolation process is $\ZZ^2$
by saying that a given point $x$ in $\ZZ^2$ is open only if the tile
$t = \phi^{-1}(x)$ is good. Now we can claim that the existence of an
edge in $\ZZ^2$ implies the existence of a path from the
representative points of the two tiles corresponding to the two end
points of the edge. We state this formally, including an observation
about the distance stretch between the two representative points.

\begin{claim}
\label{clm:coupling:nnk}
If an edge exists in the percolated mesh $\ZZ^2$ between two points
$x$ and $y$  then
\begin{enumerate}
\item There is a path between the representative points
  $\rep(\phi^{-1}(x))$ and $\rep(\phi^{-1}(y))$ of the tiles
  corresponding to $x$ and $y$ in $\NN(2,k)$ and
\item there is a constant $c_{\ktiles}$ such that
\begin{multline}
d_k(\rep(\phi^{-1}(x)), \rep(\phi^{-1}(y))) \\ \leq c_{\ktiles} \cdot
d(\rep(\phi^{-1}(x)), \rep(\phi^{-1}(y))).
\end{multline}
\end{enumerate}
\end{claim}

\begin{figure}[htbp]
\begin{center}
\input{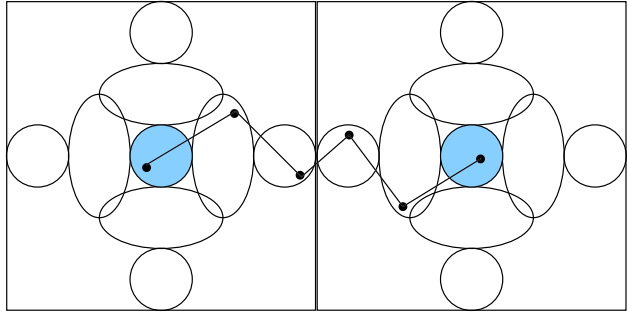}
\caption{A path between the representative points of two neighboring
  good tiles.}  
\label{fig:coupling-claim}
\end{center}
\end{figure}

\begin{proof} The proof of the claim is
  depicted in Figure~\ref{fig:coupling-claim} Clearly any circle drawn
from $\rep(t)$ that stays within $t$ contains all of $E_r$ in it by
the definition of $E_r$. Since there are at most $k/2$ points in every
good tile, hence there is an edge from $\rep(t)$ to the point
guaranteed to be contained in $E_r$, let's call it $x_r$. We do not
make any claims on where the edges established by $x_r$ to its
neighbours lie, observing only that any point that lies in $C_r$ must
have an edge to $x_r$, again by the definition of $E_r$. However, any
disc centred at a point in $C_r$ that remains within $t$ and $t_r$
must contain the left disc of its neighboring tile. Hence, if $t$ and
$t_r$ are both good then a path from $\rep(t)$ to $\rep(t_r)$
occurs. The second part of the claim is obviously true. The constant
$c_{\ktiles}$ can easily be calculated using calculus.
\end{proof}

We define $\NS(2,k)$ as the largest connected component of the graph
built on representative and relay points.  Note that unlike
$\US(2,\lambda)$ there are 8 relay points within each tile here and
the path between two representative points contains 4 relay
points. Also note that only the regions $E_l, E_r, E_t$ or $E_b$ can
share relay points. The regions $C_l, C_r, C_t$ and $C_b$ do not
intersect in any way. Each of the regions may contain more than one
point of the point process. In this case one point has to be be chosen
as a representative or relay as the case may be. This can be easily
achieved by running a simple leader election
algorithm~\cite{singh-podc:1992} between the nodes in the region which
are all connected to each other in any case. We will discuss this
issue further in Section~\ref{sec:algorithmic}.

From Claim~\ref{clm:coupling:nnk}, it is easy to deduce that if an
infinite component exists in the site percolation setting, then an
infinite component exists in $\NN(2,k)$. Hence we need to determine
for what settings of our parameters $a$ and, more importantly, $k$,
the site percolation process is supercritical.  The critical
probability for site percolation lies between 0.592 and 0.593 (see
e.g. \cite{lee-arxiv:2007}). Numerical calculations showed that the
smallest value of $k$ for which the probability of a tile being good
exceeds 0.593 is 188, and the value of $a$ for which this happens is
0.893. 

Like H\"aggstr\"om and Meester's proof for the existence of a critical
value~\cite{haggstrom-rsa:1996} and Teng and Yao's proof for the
weaker of their two upper bounds on
$k_c(2)$~\cite{teng-algorithmica:2007} our proof of
Theorem~\ref{thm:2d-bound} proceeds by constructing a coupling with a
site percolation process on $\ZZ^2$. However, our construction gives a
better upper bound than Teng and Yao's improvement of their own result
(also in~\cite{teng-algorithmica:2007}) to $k_c(2) \geq 213$ which
uses a coupling to a mixed percolation process. Hence we state this
result as a theorem:

\begin{theorem}
\label{thm:2d-bound}
For $\NN(2,k)$,
\[k_c(2) \leq 188.\]
\end{theorem}

Having described the constructions of $\US(2,\lambda)$ and $\NS(2,k)$
and having shown that there are values of the critical parameters for
which these constructions exist, let us now proceed to show that these
constructions indeed have constant stretch.

%=========================================================================
%  Constant stretch
%=========================================================================

\section{Stretch and coverage}
\label{sec:stretch}

In this section we prove that our constructions have constant stretch
with high probability when the critical parameters have high enough
values. We also show that the coverage of our constructions is very
good in a probabilistic sense.

\subsection{Constant stretch}
The argument for constant stretch of both $\NS(2,k), k \geq k_s$ and
$\US(2,\lambda), \lambda \geq \lambda_s$ follow similar lines so we
present them together. For the purposes of this section we denote the
tile lengths chosen for the two constructions as $a_u$ (= 4/3) and
$a_k$ (= 0.893). In what follows, all arguments hold for both
settings except where explicitly noted otherwise. Also in the
following we implicitly assume that $k \geq k_s$ and $\lambda \geq
\lambda_s$. 

Let us consider any two tiles $t_1$ and $t_2$ whose representative
points $\rep(t_1)$ and $\rep(t_2)$ lie $\US(2,\lambda)$ or
$\NS(2,k)$. First we relate the distance in the (unpercolated) lattice
to the euclidean distance between these two points by observing a
simple fact.
\begin{fact}
\label{fct:euclidean-lattice}
Given the constants $c_{\utiles}$ defined in
Claim~\ref{clm:coupling:udg} and $c_{\ktiles}$ defined in
Claim~\ref{clm:coupling:udg}, then for two tiles $t_1, t_2$
\[D(\phi(\rep(t_1)), \phi(\rep(t_2))) \leq \sqrt{2a} \cdot
\frac{d(\rep(t_1),\rep(t_2))}{c}\]
where $c \in \{c_{\utiles}, c_{\ktiles}\}$ and $a \in \{a_u,a_k\}$
respectively. 
\end{fact}

Fact~\ref{fct:euclidean-lattice} along with Lemma~\ref{lem:antal}
gives us the following theorem:

\begin{theorem}
\label{thm:constant-stretch}
\begin{enumerate}
\item For $\US(2,\lambda)$, with $\lambda \geq \lambdas$ there are
  constants $\alpha$ and $c_1$ depending only on $\lambda$ such that
\[\pr(d_u(x,y) > \alpha \cdot D(x,y)) < e^{-c_1 \cdot D(x,y)}.\]
\item For $\NS(2,k)$, with $k \geq 188$ there are constants $\beta$
  and $c_2$ depending only on $k$ such that
\[\pr(d_k(x,y) > \beta \cdot D(x,y)) < e^{-c_2 \cdot D(x,y)}.\]
\end{enumerate}
\end{theorem}

Theorem~\ref{thm:constant-stretch} is an existential result. In
Section~\ref{sec:algorithmic} we will show how to actually find the
constant stretch paths in a distributed way with bounded overhead. For
now we proceed to show that our constructions have good coverage.

\subsection{Coverage}

Let us consider a square region of size $\ell \times \ell$. Let us
call this $B(\ell)$. We will argue that the probability that $B(\ell)$
contains no point of $\US(2,\lambda)$ (or $\NS(2,k)$) decays
exponentially with $\ell$. As before the arguments here also apply to
both the models. We claim the following theorem:

\begin{theorem}
\label{thm:coverage}
\begin{enumerate}
\item For $\lambda \geq \lambdas$ there are constants $c_3, c_4$
  depending only on $\lambda$ such that
\[\pr[|B(\ell) \cap \US(2,\lambda)| = 0] \leq c_4 \cdot \ell^2 \cdot
  e^{-c_3 \cdot  \ell}. \] 
\item For $k \geq 188$ there are constants $c_5, c_6$ depending only
  on $k$ such that
\[\pr[|B(\ell) \cap \NS(2,k)| = 0] \leq c_5 \cdot \ell^2 \cdot e^{-c_6
  \cdot \ell}.\]
\end{enumerate}
\end{theorem}

Let us denote by $TB(\ell)$ the set of tiles fully or partially
contained in the $B(\ell)$. Let us consider the set $\phi(TB(\ell))$
i.e. the set of all points in $\ZZ^2$ which are images of the tiles in
$TB(\ell)$ under the mapping defined earlier. For $B(\ell) \cap
\US(2,\lambda)$ to be empty, each point of $\phi(TB(\ell))$ must be
outside the infinite cluster of the supercritical percolation
process. With this insight we now refer the reader to Theorems~8.18
and 8.21 of~\cite{grimmett:1999} dealing with the radius of finite
clusters in the supercritical phase. A slight modification of the
proof of Theorem 8.21 of (due to~\cite{chayes-ap:1987}) will yield the
the proof of Theorem~\ref{thm:coverage}. The details are tedious and
do not add anything to the proof described in~\cite{grimmett:1999} so
we omit them here. Theorem~\ref{thm:coverage} yields the following
simple corollary

\begin{corollary}
\begin{enumerate}
\item There is a constant $c_7$ such that for $\ell \geq c_7 \log n$
\[\pr[|B(\ell) \cap \US(2,\lambda)| = 0] < \frac{1}{n}.\]
\item There is a constant $c_8$ such that for $\ell \geq c_8 \log n$
\[\pr[|B(\ell) \cap \NS(2,k)| = 0] < \frac{1}{n}.\]
\end{enumerate}
\end{corollary}

The constants $c_7$ and $c_8$ may be larger than what the network
requires for its sensing function. It seems intuitive that adding more
nodes should decrease the values of these constants, but the statement
of Theorem~\ref{thm:coverage} does not seem to provide this. However
we claim this intuition is indeed satisfied for $\US(2,\lambda)$ and
$\NS(2,k)$. In order to use Theorem~\ref{thm:coverage} to provide a
guarantee of desired coverage, we argue that the exponential decay
indeed grows sharper when the density $\lambda$ increases. This is
true for both $\US(2,\lambda)$ and $\NS(2,k)$. 

For the case of $\US(2,\lambda)$ it is not hard to see, since an
increase in $\lambda$ directly leads to an increase in the probability
of a tile being good. This, through the coupling, increases the
probability of the corresponding site in $\ZZ^2$ being open. In the
site percolation setting the probability of a site being part of the
infinite cluster, $\theta(p)$ is known to increase montonically with
the probability $p$ of sites being open. Our claim follows because the
increase in $\theta(p)$ is centrally involved in the (omitted) proof
of Theorem~\ref{thm:coverage}. The claim has a slightly subtler
provenance for $\NS(2,k)$. Essentially the argument is that if we fix
some value $k \geq k_s$, increasing the density $\lambda$ allows us to
use tiles of smaller side length and still achieve the desired
probability ($>p_c$) of a tile being good. This implies that the
number of tiles within a region increases, and hence the exponential
decrease becomes sharper.

%=========================================================================
%  Algorithmic issues
%=========================================================================

\section{Algorithmic issues}
\label{sec:algorithmic}

We now focus on the algorithmic issues involved in building
$\US(2,\lambda)$ and $\NN(2,k)$ and routing packets in them once they
are built. 

\subsection{Forming the networks}

After the nodes are laid out in their positions they have to undertake
four basic steps. Firstly, they have to identify which tile they
belong to. This involves using their location information (assumed to
be of the form $(x,y) \in \RR^2$) and the value of the tile width
(denoted $a_u$ for $\US(2,\lambda)$ and $a_k$ for $\NS(2,k)$)
programmed into the nodes. In the second step each node determines
whether it belongs to one of the special regions within the tile as
described in Section~\ref{sec:coupling}. In the third step all the
nodes within a region communicate to elect a leader who is then
designated as the representative point of the tile or a relay point,
as applicable. In the fourth step the elected points of each region
form connections with the leaders of their neighbouring regions. See
Figure~\ref{fig:formation} for a formal statement of the algorithm for
building $\US(2,\lambda)$. The algorithm for $\NS(2,k)$ is very
similar.

\begin{figure}[ht]
\begin{center}
\fbox{
\begin{minipage}{0.9\columnwidth}
\noindent{Algorithm {\sf construct}($\US(2,\lambda), a$)}
\begin{enumerate}
\item At each node $v$ do
\begin{enumerate}
\item Determine $(\loc_v(x),\loc_v(y))$.
\item compute $\id_v(x) = \loc_v(x)/a_u$ 
\item compute $\id_v(y) = \loc_v(y)/a_u$
\item compute $\region_v$
\end{enumerate}
\item For each tile $t$ and each region $r \in \{C_0(t), E_r(t),
  E_l(t), E_t(t), E_b(t)\}$ do
\begin{enumerate}
\item Build $S(r,t) = \{v \in t \ |\ \region_v =  r\}$.
\item $\rep(t) \leftarrow$ {\sf electLeader}$(S(C_0,t))$
\item For $r \in \{ E_r(t),  E_l(t), E_t(t), E_b(t)\}$\\ 
$\relay(t,r) \leftarrow$ {\sf electLeader}$(S(r,t))$
\end{enumerate}
\item For each tile $t$ with neighbours $t_l, t_r, t_t, t_b$ do 
\begin{enumerate}
\item For each $r \in \{ E_r(t),  E_l(t), E_t(t), E_b(t)\}$\\
{\sf connect}$(\rep(t), \relay(t,r))$.
\item {\sf connect}$(\relay(t,E_r(t)), \relay(t,E_l(t_r))$.
\item {\sf connect}$(\relay(t,E_l(t)), \relay(t,E_r(t_l))$.
\item {\sf connect}$(\relay(t,E_t(t)), \relay(t,E_b(t_t))$.
\item {\sf connect}$(\relay(t,E_b(t)), \relay(t,E_t(t_b))$.
\end{enumerate}
\end{enumerate}
\end{minipage}
}
\caption{Building $\US(2,\lambda)$.}
\label{fig:formation}
\end{center}
\end{figure}

The function {\sf electLeader} can be realized using any distributed
leader election algorithm on a complete graph topology since all the
nodes within a region can talk to each other (see
e.g.~\cite{singh-podc:1992}. The function named {\sf connect} is
simply a handshake between the two nodes mentioned. Once the calls to
this function are over the set up phase is completed.

Note that the algorithm of Figure~\ref{fig:formation} will not just
form the largest component but will also form other small
components. It is possible to detect how large a given component is by
attempting to send packets to distant nodes. The nodes of a small
component can then turn themselves off if they realize they are not
part of $\US(2,\lambda)$. Detecting connectivity is an area of
research in itself so we do not address the issues here, refering the
reader to some recent work in this area~\cite{jorgic-icccn:2007}.

\subsection{The routing algorithm}

\begin{figure}[htbp]
\begin{center}
\input{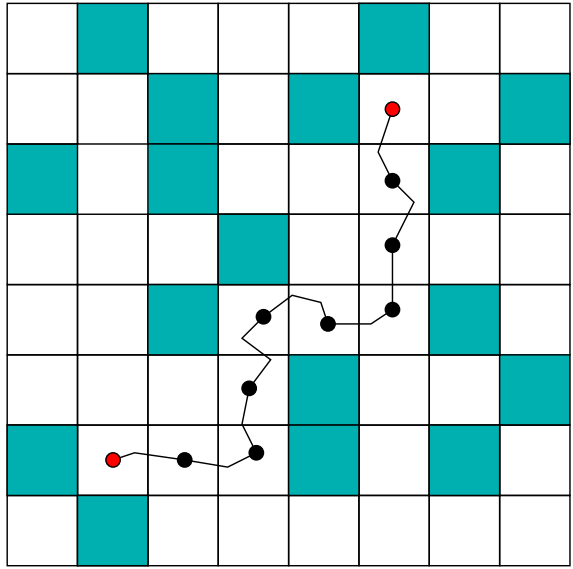}
\caption{The path between two representative points mimics the path in
  $\ZZ^2$ using relay points to realize edges.} 
\label{fig:metric-path}
\end{center}
\end{figure}

For routing purposes the representative points of a tile act as if
they are open lattice points in $\ZZ^2$. They use relay points to send
packets to the representative points of their neighbouring good tiles
(see Figure~\ref{fig:metric-path}) hence realizing open edges in
$\ZZ^2$. With this simple idea in place, we can just plug in any
algorithm which performs routing in the percolated mesh.

We rely on the algorithm for efficient distributed routing in the
giant component of a percolated mesh given by Angel
et. al.~\cite{angel-podc:2005}. Their algorithm proceeds by trying to
follow a shortest path from source to destination. If at any point the
path is broken (i.e. one of the nodes is closed) they try to find the
next node along the path that is open by performing a distributed BFS
from the current location of the packet. For our purposes we assume
that the canonical shortest path between any two nodes $(x_1,y_1),
(x_2,y_2) \in \ZZ^2$ is the $x-y$ path: the path that proceeds to
first fix the $x$ coordinate then the $y$ coordinate i.e. $(x_1,y_1)
\rightarrow (x_2,y_1) \rightarrow (x_2,y_2)$. We describe the routing
algorithm in Figure~\ref{fig:routing}.

\begin{figure}[ht]
\begin{center}
\fbox{
\begin{minipage}{0.9\columnwidth}
\noindent{Algorithm {\sf routing}($\rep(t_1), \rep(t_2)$)}
\begin{enumerate}
\item $\id_s \leftarrow \id_{\rep(t_1)}$.
\item $\id_t \leftarrow \id_{\rep(t_2)}$.
\item $curr \leftarrow \id_s$.
\item While $curr \ne \id_t$ do 
\begin{enumerate}
\item $next \leftarrow$ {\sf computeNext}($curr, \id_t$)
\item if {\sf isOpen}($next$)
\begin{enumerate}
\item  {\sf sendTo}($next$)
\item  $curr \leftarrow next$
\end{enumerate}
else 
\begin{enumerate}
\item run {\sf distBFS}($curr, \id_t$) until
 $v$ lying on the $x-y$ path from $curr$ to $\id_t$
is found.
\item  {\sf sendToNode}($v$).
\item  $curr \leftarrow v$.
\end{enumerate}
\end{enumerate}
\end{enumerate}
\end{minipage}
}
\caption{Routing packets between the representative points of two
  tiles.}
\label{fig:routing}
\end{center}
\end{figure}

The function {\sf computeNext} mentioned in Figure~\ref{fig:routing}
finds the next node along the $x-y$ path. The function {\sf isOpen}
involves checking if the next tile along the path is good or not. This
can be done by asking the relevant relay if it has a neighbour in the
target tile. The function {\sf sendTo} uses the relays to pass the
packet to the representative node of the next tile which then
continues the routing process. This is the simple part of the
algorithm. If the target tile is not good a BFS is launched to find
the next good tile along the $x-y$ path. This is a distributed
algorithm that requires nodes to be probed as the search
proceeds. Finally when it finds the destination it has to also report
the path back to the node that launched the search. Once this path is
known the function {\sf sendToNode} sends the packet along this path
to the discovered node. We refer the reader to Angel
et. al.~\cite{angel-podc:2005} for a proof that the expected number of
probes required for this algorithm is at most a constant times the
length of the shortest path.

%=========================================================================
%  Conclusion
%=========================================================================

\section{Conclusion}
\label{sec:conclusion}

In this paper we have shown that it is possible to construct sparse
power-efficient wireless ad hoc sensor networks with good coverage if
we are willing to accept a certain level of redundancy in the
system. Ideas from percolation theory have been used to demonstrate
that the infinite cluster of two kinds of geometric random graphs
contains a good subgraph with the properties that we seek and that
this subgraph can be built efficiently using only local
information. It is our conjecture that the subgraphs we build should
exist whenever an infinite cluster exists in the geometric random
graphs we study. One major direction for future research involves
resolving this conjecture one way or the other. Even if this
conjecture is not true, it should be possible to bring the values of
$k_s$ and $\lambda_s$ closer to the critical values $\lambda^{(2)}_c$
and $k_c(2)$.

%=========================================================================
%  Bibliography
%=========================================================================

% \newpage
\bibliographystyle{abbrv}

\end{document}